# (PREPRINT) ROSETLineBot: One-Wheel-Drive Low-Cost Power Line Inspection Robot Design and Control

A. Tarık Zengin[1], Gokhan Erdemir[2], T. Cetin Akinci[3], F. Anil Selcuk[1,4], M. Nizamettin Erduran[1], S. Serhat Seker[3]

[1] Istanbul Sabahattin Zaim University, Dep. of Computer Engineering, Istanbul, Turkey; {tarik.zengin, nizamettin.erduran}@izu.edu.tr
[2] Istanbul Sabahattin Zaim University, Dep. of Electrical and Electronics Engineering, Istanbul, Turkey; gokhan.erdemir@izu.edu.tr
[3] Istanbul Technical University, Department of Electrical Engineering, Istanbul, Turkey; {akincit, sekers}@itu.edu.tr
[4] Marmara University, Department of Mechatronics Engineering, Istanbul, Turkey; anilselcuk@marun.edu.tr
[2] Correspondence author

**Abstract:** Continuous operation of the electrical transmission lines is of great importance for today's electricity-dependent world. It is crucial to detect and diagnose a possible problem on the lines before that occurs. In general, maintenance, repair and error detection operations of the electrical transmission lines are performed by humans periodically. However, these difficult tasks contain high-risks in terms of occupational health and safety. Moreover, it is not possible to continuously inspect these lines which extend over long distances by maintenance and repair personnel. For all these reasons, it is very convenient to use robotic systems for inspection on power transmission lines. By evaluating the data which are provided from sensors and cameras of robotic systems, it would be possible to take precautions before an error occurs. Robotic systems that can move continuously along the line would be able to gather continuous and uninterrupted information. Thus, potential problems can be detected and identified in advance. In this study, a single wheel drive low-cost mobile robot is designed and controlled which can work and move on the electrical transmission lines. Because of the modular structure of the designed robot, different types of sensors can be integrated into the robot, easily. The designed low-cost robotic system will cause minimum losses in case of possible breakage.

## 1. Introduction

Energy dependency of today's world obliges energy transmission lines to more productive and uninterrupted operation without any problem[1]. Energy transmission lines (systems) face many challenges due to the growing need for sustainable energy worldwide [1-5]. It is important to do maintenance, repair, breakdown, leak detection, and etc. operations of the power lines, rapidly [1-26]. Although the operations to be carried out on power transmission lines are often performed by humans, it can be fully or partially automated by using robotic systems with direct or teleoperated (remote control) robot systems [1-10]. There are various risks to humans who will work in the maintenance of power lines and transformers in terms of occupational health and safety. The maintenance of power lines contains various risks for humans who will work for maintenance operations in terms of occupational health and safety. Using robotic systems on the power lines can increase productivity, reduce labor costs and, most importantly, eliminate operational risks that threaten human life[1]. Although, development and implementation of the robotic applications for various tasks on the power lines is still so challenging and popular research area in robotics science [1-5]. The specifications and characteristics of the robotic systems which will be used on the power line inspection depend on the tasks to be performed on the power lines[1-10]. Research on energy transmission systems in robotics generally focus on high voltage lines [10-15, 21-26]. It is a suitable research area for the use of robotic systems due to the fact that high-voltage lines are far from the urban areas, being high above the ground, and carrying a large number of risks for the maintenance personnel[5-12].

In literature, robotic systems used in high voltage lines vary. Generally, wheeled systems that are seated on the cable and grasp the cable tightly are preferred [8]. Some of the developed robotic systems for power line inspections are shown in Figure 1. The "Expliner©" robot which is developed by HiBot has been used in Japan, South Korea, the Netherlands and the United States [27]. The robot, developed by the EPRI (Electric Power Research Institute), is actively used in high-voltage lines between cities or states, especially in the United States [28].

Medium voltage lines are also another research area for robotic systems [1, 3, 5-10]. However, some problems which are listed below in medium voltage lines make it difficult to use robotic systems in these lines. General problems in medium voltage line; (i) the absence of a certain standard in the types of electricity pole, (ii) using different types of materials(cable, bowl, connection apparatus, etc.), (iii) they are densely located within the urban areas, and (iv) they have different connection types. Because of the mentioned problems, it is more appropriate to use semi-autonomous or teleoperated robotics systems instead of fully autonomous systems in medium voltage lines [1-5].

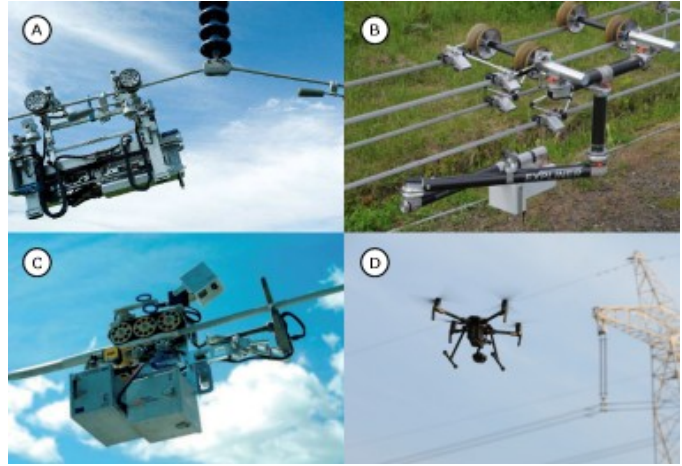

**Figure 1.** (a) LineScout robot [18], (b) HiBot "Expliner©" [27], (c) LineROVer [1], and (d) DJI – M200 – Powerline Inspection Tool[29].

In this study, the design, and control of a one-wheel-drive low-cost power line ınspection robot which is named "ROSETLineBot" is proposed. In section 2, the structure, components and design of ROSETLineBot are explained. Then the components of ROSETLineBot are described in the same section. The system integration and testing results are presented in section 3. Finally, our conclusions are presented in section IV.

## 2. The structure of ROSETLineBot

In this section, 3D design steps, wheel kinematics, rothe robot electronics and control design of the robot are given in detail.

## 2.1. 3D Model of ROSETLineBot

In the prototyping stage, SOLIDWORKS software was used in the design of the ROSETLineBot. In the prototyping phase, PLA filament was used for 3D printing of the robot. Each part of the robot was printing, separately. And then, each component was combined by using special fasteners.

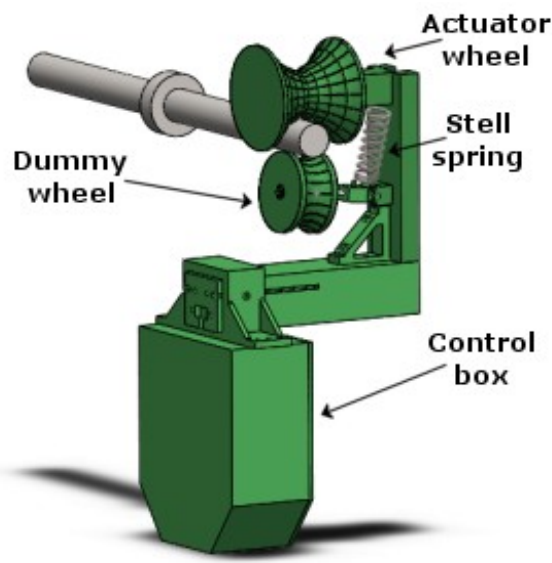

**Figure 2.** 3D model of ROSETLineBot.

The structure of the robot mechanism consists of the mainframe, actuator wheel, dummy wheel, and control box. The dummy wheel is mounted to the mainframe with the steel spring. Shafts produced from ASP23 high-speed steel have been utilized in the special dimensions determined for strengthening the main body and the control box.

The robot moves on the test line by DC motor which is connected to the actuator wheel. The actuator and dummy wheels grasp the power line by means of pulling force of the steel spring. While the actuator wheel moves only in its own axis, the dummy wheel can move both in its own axis and in the vertical axis by force of the steel spring. When the robot faces an obstacle while it is moving on the line, the spring opens when the actuator wheel is trying to pass the obstacle, the robot gains suitable grip space with the expansion of the steel spring on the vertical axis. Thus, the obstacles can be easily crossed. After passing the obstacle, the spring closes and the dummy wheel takes its actual position and keeps going to its movement on the power line.

## 2.2. Wheel Kinematics

In this section, we consider only a simplified case of the robot movement for which the longitudinal slip between the wheels and the power line can be neglected. In [30], velocities and forces on one wheel are described in detail. Forces and velocities of a wheel which described in [30] are shown in Figure 3. A quarter vehicle model was used to calculate velocity of the robot in this study. Equation (1) was used to calculate the velocity of the robot.

$$v = \frac{C_{nt} \times C}{C_{pr} \times n \times d_t}$$

(1)

In Equation (1), $C_{nt}$ is the encoder count, C is the circumference of the wheel, $C_{pr}$ is the counts per rotation of encoder, n is the gear ratio of motor reductor and $d_t$ is the count period.

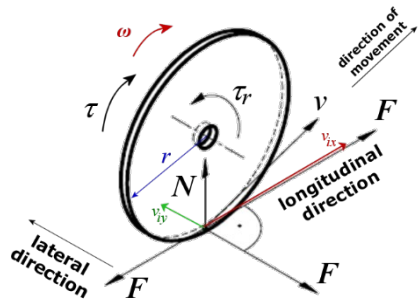

**Figure 3.** Wheel description for one-wheel-drive system [30].

## 2.3. Electronics Circuit Design

Electronics components of robot were chosen it to be affordable as it is aimed at one of the main objectives. Block diagram of the electronics circuit is shown in Figure 4. An ATMega2560 microcontroller together with a L298 motor driver were used as the main control unit. It got the velocity feedback from the quadrature encoder integrated with the DC motor. The encoder had 34:1 gear ratio and 48 CPR, hence the total counts per revolution was 1632. That brought precision to the measurement since the encoder is quite sensitive.

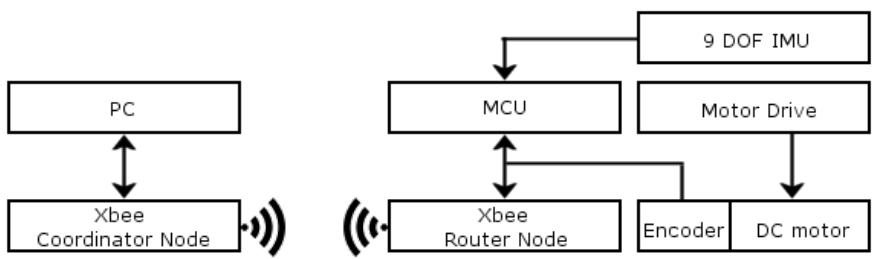

**Figure 4.** Block diagram of the electronics circuit.

Besides the quadrature encoder, a 9DOF IMU was placed onto the robot in order to acquire orientation and acceleration. Additionally, an XBee Pro module was used for remote operation and parameter setting. Although an S2B model having a 120m range was used in this experiment, it's easy to change it to a long range model since they're pin compatible.

## 2.4. Control Algorithm

PID control which is shown in Equation (2) was used to control the velocity of the robot. The feedback was obtained by data that was gathering from the motor encoder. Encoder values were converted to the velocity values by using Equation (1).

$$u_{(t)} = K_p e_{(t)} + K_i \int_0^t e_{(t)} dt + K_d \frac{d}{dt} e_{(t)} \qquad (2)$$

In Equation (2), $u_{(t)}$ is the control signal. $K_p$, $K_i$, and $K_d$ are proportional, integral and derivative coefficients, respectively. $e_{(t)}$ is the error between actual and desired speed. In the PID control block, its coefficients were selected as . $K_p = 30$, $K_i = 1$, and $K_d = 0.1$ by experimentally. The block diagram of the control algorithm is shown in Figure 5.

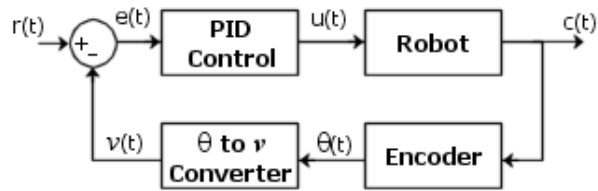

**Figure 5.** Block diagram of the control algorithm.

## 3. Experimental Studies

In this section, experimental results are shown graphically. Pictures of ROSETLineBot on the test line from different perspectives is shown in Figure 6. In the experiment, measured velocity, pitch, roll, and yaw values are shown in Figure 7 and 8 when reference velocity set as $u_{(t)} = 20 \, cm/sec$. According to test results which are shown in figures 7 and 8, PID control shows satisfying performance for the power line inspection robot. The experimental setup was conducted with a 25mm diameter sagged actual power line cable in the laboratory. The objective was to maintain the velocity of robot on a sagged line. Experimental results show the robot climbing the sagged cable. Figure 7 shows that the controller maintained the velocity by increasing the control signal during the climbing and achieved an almost-stable ride. Figure 8 shows the orientation during ride. The oscillation was kept in ±10 degrees.

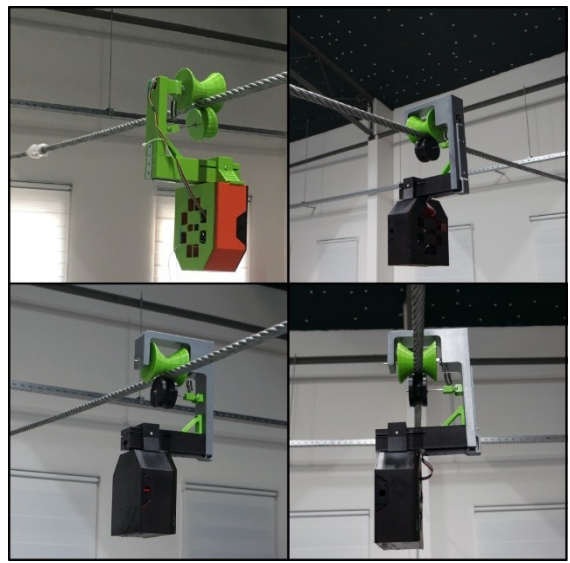

**Figure 6.** Pictures of ROSETLineBot on the test line from different perspectives.

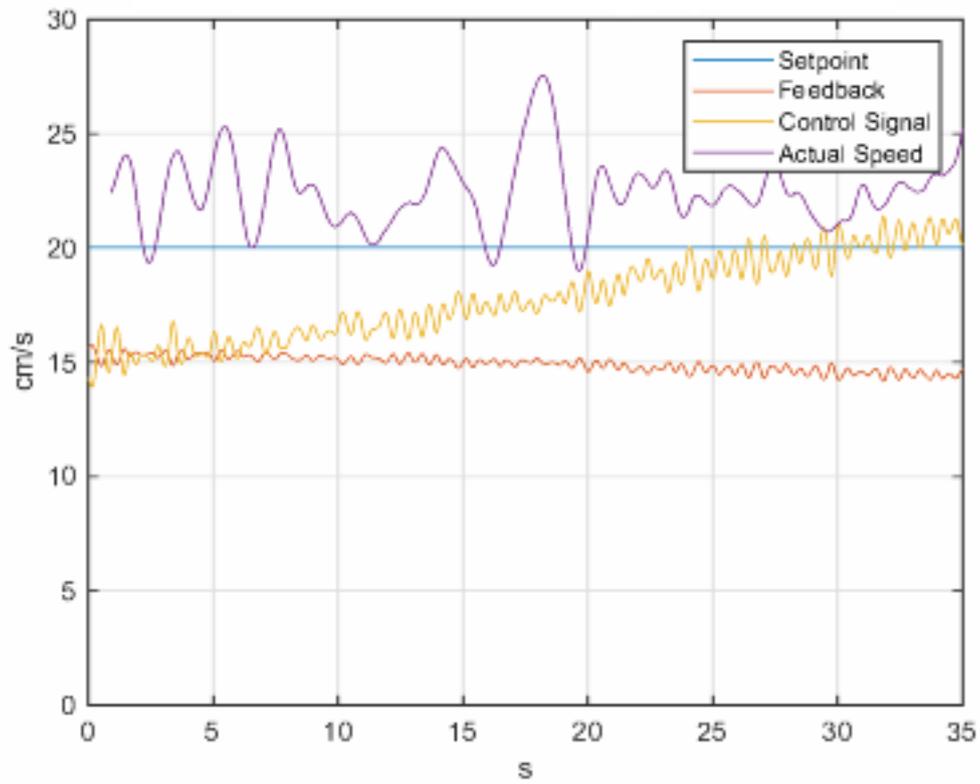

**Figure 7.** Speed measurement on a sagged power line during speed control test.

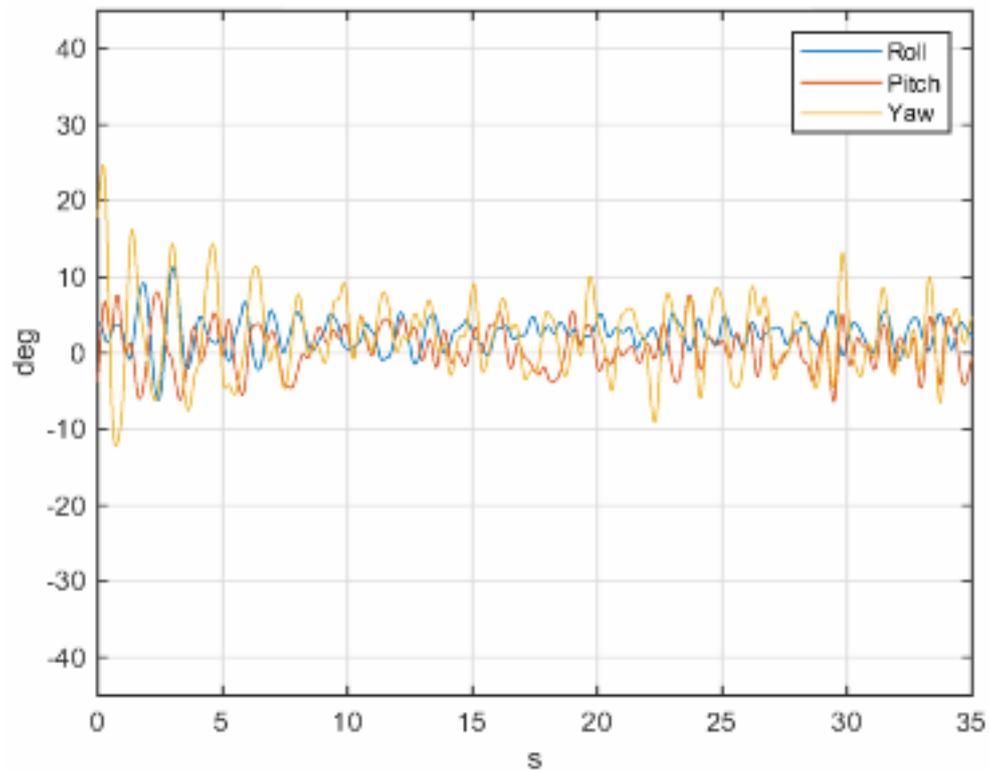

**Figure 8.** Roll, pieth, and yaw angles of ROSETLineBot during speed control test.

## 4. Conclusion

The main purpose of the study is to examine the feasibility of error (problem) detection and identification operations which can be performed by robotic systems on the power lines. Operations on the power line contain high-risks in terms of occupational health and safety for humans. These risks sometimes lead to undesirable results. To prevent these deadly results, robotic systems can be used for maintenance, repair and error detection operations of the electrical transmission lines. In this study, a single-wheel-drive low-cost fully 3D printed mobile robot is designed and controlled for inspection of the power transmission lines. In the experimental study which is presented in section 3 in detail, the velocity control of the robot was performed by using PID control. It was observed that the robot was able to strongly grasp the power wire and smoothly move on it. A sensor fusion control method was remained as the future work.